# Additive Manufacturing and Characterization of Architectured Cement-based Materials via X-ray Micro-Computed Tomography

Mohamadreza Moini[1*], Jan Olek[1], Bryan Magee[2], Pablo Zavattieri[1], Jeffrey Youngblood[1]

[1] Purdue University, West Lafayette IN 47909, USA
[2] Ulster University, Newtownabbey, BT370QB, UK
*corresponding author: mmoini@purdue.edu

**Abstract.** There is an increasing interest in the fabrication of cement-based materials via additive manufacturing (AM) techniques. However, the processing-induced heterogeneities and interfaces represent a major challenge. The role of processing in creating interfaces and their characteristics requires understanding of the microstructure of 3D-printed hardened cement paste (hcp). This work investigates the microstructural features of architectured cement-based materials, including processing-induced heterogeneous patterns, interfacial regions (IRs), and pore network distributions with respect to the architectural pattern. A 3D printer was modified and merged with an extrusion system and specimens were 3D-printed using a layer-wise direct ink writing (DIW) process capable of fabrication of 'lamellar' architectures of materials. A lab-based X-ray microscope (XRM) was used to perform X-ray micro-computed tomography (micro-CT) evaluations to explore the microstructural characteristics of 3-day old intact (i.e. not tested) 3D printed and cast specimens at two levels of magnification: 0.4X and 4X. CT scans of printed specimen revealed a patterned pore network and several microstructural features, including: a) macropores (visible during printing), b) micropores at interfacial regions (IRs), c) accumulation of anhydrous cement particles near macropores, and d) rearrangement of filaments away from their designed toolpath. In comparison, microstructural investigation of cast specimen at 4X scan revealed randomly distributed pores with no connectivity throughout the specimen. The aptitude of micro-CT as a non-destructive technique for microstructural characterization of architectured cement-based materials is discussed. The role of processing to induce and to pattern heterogeneities such as IRs in materials is demonstrated and the role of architecture in controlling such heterogeneities and their directionality through the interface is discussed.

**Keywords:** 3D-printing, Cement Paste, Micro-CT, Interfacial Region (IR)

## 1    Introduction

Properties of hardened cement paste (hcp) are influenced by its microstructure and the way in which the material is cast and placed [1]. Specifically, characteristics of the pore network (i.e., the size and distribution of internal flaws), the morphology of microstructural components, and heterogeneities in the microstructure, all affect mechanical properties of hardened cement-based materials [1,2]. Over the last decades, the use of advanced characterization techniques, such as scanning electron microscopy (SEM), has significantly advanced the understanding of microstructure of cement-based materials. However, the effectiveness of SEM is limited in terms of obtaining three-dimensional information about connectivity and size distribution of pore network [1,3].

X-ray micro-Computed Tomography (micro-CT) has been previously applied to cement-based mortar and concrete materials to characterize pore network [4] and microstructure [5] and their relationships with a variety of key properties including: fracture properties [5,6,7], damage mechanisms [2,8,9], mass transport [10], and evolution of cement hydration [11]. Micro-CT is a non-destructive technique that captures three-dimensional images of materials without the need for destructive preparation processes such as drying, surface treatments, and vacuuming; all of which are commonly required when preparing specimens for typical microstructural characterization techniques such as MIP, gas sorption, and SEM [1]. As such, the use of micro-CT for microstructural characterization of 3D-printed hcp is advantageous as it does not alter the microstructure while allowing for imaging of elements as large as 10s of millimeters.

During this research, a laboratory-based (i.e. not requiring synchrotron facilities) X-ray microscope was used to explore the microstructural characteristics of intact (i.e. not tested) printed and cast 3-day old cement paste specimens at two magnifications: 0.4X and 4X corresponding, respectively, to the resolution of $32.24\ \mu m$ and $4.04\ \mu m$.

3D-printing via direct-ink-writing (DIW) of colloids, slurries, and pastes allows for control of the architecture of the element and can give rise to a variety of microstructural features [12,13,14]. In this work, four microstructural features observed in 3D-printed lamellar architecture are discussed. All four features were qualitatively detected at 0.4X scans, and further verified at 4X scan. Finally, the microstructure of the printed specimen was compared to a microstructure of conventionally cast specimen.

## 2    Methods

### 2.1    3D-Printing Setup

To establish a 3D printer capable of printing cement paste via DIW techniques, two separate units were combined. A gantry-based 3D printer (Ultimaker 2 Extended+ used for printing thermoplastic materials) was merged with a stepper motor-driven extrusion system (Structur3d Discov3ry Paste Extruder) to serve as a paste extruder (see Fig. 1(a)). These two units were connected with tube and luer locks. The resulting system is capable of printing pastes with high yield stress and viscosity and can fabricate elements at prototyping scale (mm). The extruder is capable of applying desirable displacements (extrusion rates) via mounted 75 mL ink-charged syringes as depicted in Fig. 1(b). The 3D printer was modified to allow for mounting of a nozzle holder assembly on the printer gantry rods (Fig. 1(c)). The nozzle holder is consisted of two light-weight aluminum parts designed and custom-fabricated specifically for this printer. To continuously feed the paste from the syringe to the printer nozzle, a polyethylene tube for paste delivery (with an internal diameter of 4.3 mm and 450 mm long) was passed through the nozzle holder assembly. A female luer lock was inserted into the nozzle holder to connect the tube to the nozzle (Fig. 1(d)). To connect the other side of the tube, onto the syringe a male luer lock (Fig. 1(e)). The luer locks and the standard Nordsen nozzle (gauge 15 with 1.36 I.D.) are shown Fig. 1(f).

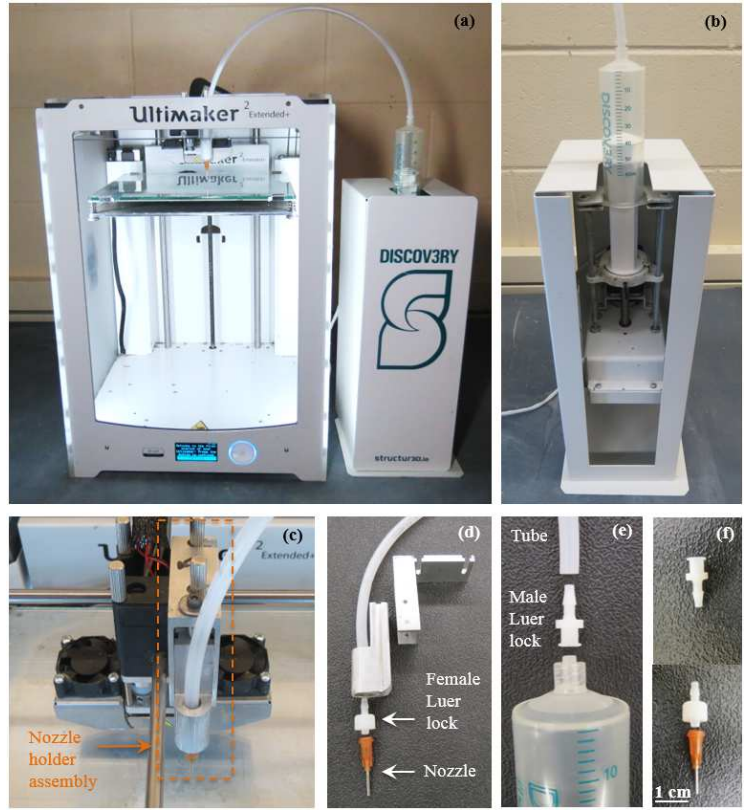

**Fig. 1.** 3D printer setup: **a)** Gantry 3D printer (Ultimaker 2 Extended +) and stepper motor-based extrusion (Discov3ry) system; **b)** Syringe and plunger mounted onto the stepper motor system; **c)** Nozzle holder assembly mounted onto the 3D printer gantry guide rods; **d)** Nozzle holder assembly and the female luer lock used to connect the tube to the nozzle; **e)** male luer lock used to connect syringe to the tube; **f)** nozzle and luer locks shown separately

## 2.2    Ink Design, Mixing Procedure, and Curing

An iterative mix design process was employed to identify cement paste inks with flow properties (i.e., yield stress, and viscosity) suitable for the DIW process. A combination of high-range-water-reducing-admixture (HRWRA) and viscosity modifying admixture (VMA) are widely used in self-consolidating concrete to avoid opposing effects of segregation at rest and high fluidity during pumping [15]. In DIW process of cement paste, it is also critical to achieve yields stresses that are high enough to allow the materials to be self-supporting (shape-holding). In the process of DIW of ceramic slurries (on the mm scale) the increase in the content of solids results in both, the increase in yield stress and in viscosity [16]. Moreover, the yield stress of cement paste is known to decrease with the addition of superplasticizers and the apparent viscosity is typically known to increase with the addition of VMAs [15]. In this work, the successful ink was designed with yield high enough for shape-holding and viscosity suitable for extrusion. A low water to cement ratio of 0.27, corresponding to a solid content of 53% (by mass), was used in this work. This percentage of solid content is found to provide suitable in ceramic paste as well [16]. In the case of cement paste, such low water to cement ratio can produce very stiff mixtures that can experience bleeding while undergoing extrusion. As such, a HRWRA was used to lower yield stress (and to ensure extrudability) and VMA was used to reduce bleeding. The use of VMA is highly desirable for DIW process. Specifically, it enhances the stability of the ink during extrusion and that of the specimen upon deposition [15].

The final ink used in this work consisted of the sub 150 µm fraction of commercially available Type I cement (ASTM C150 [17]) obtained from Buzzi Unicem, USA; the deionized water; HRWRA (MasterGlenium 7700), and VMA (MasterMatrix 362). Both chemical admixtures met the requirements of the ASTM C494 [18]. Optimal dosages were established based on findings of a related previous study [19]. For each 250.0 g of cement, the mix contained 65.2, 1.1 and 3.0 grams of deionized water, HRWRA and VMA, respectively. HRWRA and VMA were added to the water consecutively and stirred until they could not be visually observed. The liquid phase was then added to cement. A Twister Evolution Venturi vacuum mixer was used for mixing the paste to eliminate entrapped air as the presence of the air bubbles will degrade the quality of the ink. The pre-mixing mode of the mixer was used during the first 25 seconds of the mixing to process the paste at slow speed while subjected to a 70% of vacuum level provided by the mixer. This was followed by mixing at 400 rpm for 90 seconds at 70% vacuum level. The paste was mixed for a second time at 400 rpm for 90 seconds at 100% vacuum level of the mixer. The paste was then loaded into the syringe. The syringe was then outfitted with the plunger and mounted on the extruder as depicted in Fig. 1(b). The mixing process was performed within 5 min after combining cement and liquid. The specimens were cast and printed in lab environment at $18 \pm 3$ °C and $45 \pm 5$ % relative humidity. Immediately after printing (or casting), the specimens were placed in a sealed curing box which maintained constant relative humidity of $93 \pm 2$ % (by using saturated solution of potassium nitrate). The box was kept at constant temperature of $18 \pm 3$°C.

### 2.3    Slicing and Design

In order to generate a toolpath required for lamellar architecture, a commercially available slicer (Simplify3D) was used to generate the G-code commands. A cubical 3D object was introduced to the slicer and geometrical parameters were assigned to it to achieve desired printing path and architecture. The G-code command included 5 axis of control: Point cloud coordinates (X, Y, Z axis) to control the movements of the nozzle and the bed; extrusion (E axis) to control the amount of extrusion relative to the nozzle movement; and printing speed (F axis) to control the speed of the nozzle movement. E and F axis were controlled via an extrusion rate multiplier and printing speed in the slicer. Several other printing parameters, including the location, the amount, and the speed of retraction, were also scripted to the G-code in the slicer. To generate the tool path, a 25x25x25 mm cube was introduced to the slicer and a continuous printing path in each layer was designed to create lamellar architectures (Fig. 2(a)). To achieve a solid specimen, 100% infill was used for lamellar architectures (Fig. 2(b)). A layer height (filament height) of 1.00 mm and the internal diameter of the nozzle of 1.36 mm were specified in the slicer. The specified printing speed of 250 mm/min employed, resulting in a speed of 87 seconds/layer. These printing and geometrical parameters were established by trial and error to obtain a suitable print quality (i.e., filament width and height close to that specified in the slicer). A schematic cross-section of the lamellar architecture considered and a resulting specimen are shown in Figs. 2(c) and 2(d).

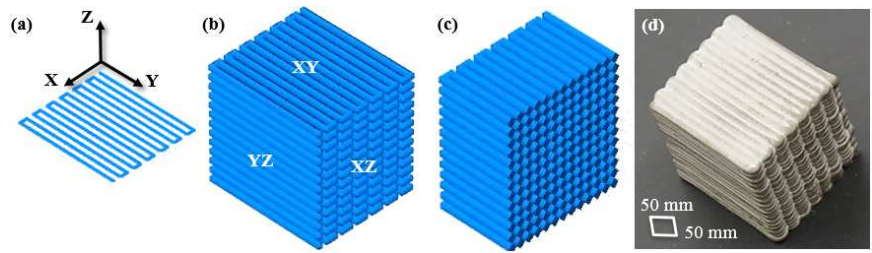

**Fig. 2.** Schematics of lamellar architectures: **a)** Printing path of individual layers; **b)** Printing path of a cube specimen and; **c)** Cross-section of the specimen with lamellar architecture; **d)** 3D-printed lamellar architecture cube via DIW

## 2.4    Micro-CT and Scanning Specimens

X-ray micro-computed tomography (micro-CT) is an imaging technique that involves the recording of series of 2D radiographs (images), taken at various angles around a rotating object, to mathematically reconstruct a spatial map and digitally render the entire volume (i.e., three-dimensional appearance) of an object [2]. Resulting 3D renditions are typically presented as a series of 2D (i.e., sliced images) with intensities corresponding to X-ray absorption and material density at each voxel [20]. The resulting variations in intensity allow for identification of various phases and features of the microstructure and their 3D distribution.

Conventionally, X-ray microscopes employed for micro-CT characterization technique, use a flat panel detector and thus rely on single-step (i.e., geometric) magnification. As a result, resolution degrades with increasing sample size and working distance. In this study, an X-ray microscope (XRM), Zeiss Xradia 510 Versa was utilized, which allows of an increase in the resolution of scans through dual-stage magnification process. In the first stage, the field of view (FOV) desirable to scan the entire volume of the specimen was established via geometric magnification process, which involved setting distances between the source, detector, and specimen (as in conventional micro-CTs). In a second stage, additional optical magnification was enabled at the detector system through objective lenses. The detector is equipped with scintillator and objective lens which converts X-rays to light rays and thus allows for optical magnification and higher resolution. The initial (i.e. the 0.4X) scan allowed a large FOV and thus facilitated to scan the entire specimens (32.24 µm pixel size). This was followed by a 4X scan, allowing higher resolution (4.04 µm pixel size) at regions of interest (ROI). A beam energy of 150 KeV, a power of 10 W, exposure times of 0.94 second and 4 seconds, and full 360° rotation were used for 0.4X and 4X scans of printed specimen and a beam energy of 140 KeV, a power of 9 W, exposure times of 1 second, and full 360° rotation were used for 0.4X scans of cast specimen respectively. Dragonfly software was used for post-processing of the data. One cast and one 3D-printed hcp cubes (25x25x25 mm ± 1 mm) were used in this experiment.

## 3    Results and Discussion

### 3.1    3D-Printed Lamellar Architecture Micro-CT (0.4X and 4X scans)

**0.4X Scan.** The 0.4X magnification CT scan of the intact specimen revealed the presence of four microstructural features: macropores, micropores, rearrangement of filaments and accumulation of anhydrous cement grains. These features are illustrated in Figs. 3 (a), 3(b) and 3(c), for XZ, YZ, XY planes respectively, and in Fig. 3(d) for the 3D rendition of the interior of the lamellar architecture specimen. In micro-CT images of hcp, darker intensities represent pores filled with air or water, with greyscale intensities corresponding to hydrated cement paste products and brighter regions corresponding to anhydrous cement grains [1]. The previously mentioned four microstructural features are analyzed in more details below:

*Macropores.* As seen in Figs. 3(a) and 3(d), there are several regions that contain large pores (macropores) located between adjacent filaments of the same layer. These pores are aligned along the filament in the direction of the Y axis as can be seen in the regions outlined by white rectangles in Figs. 3(a) and 3(d). They are typically wider than 100 µm and range in length from few to 10s of millimeters. These macropores are the result of variability in the width of the filament as it is being extruded from the tip of the nozzle. That variability is, in turn, likely the result of inconsistencies in the local properties of ink under extrusion.

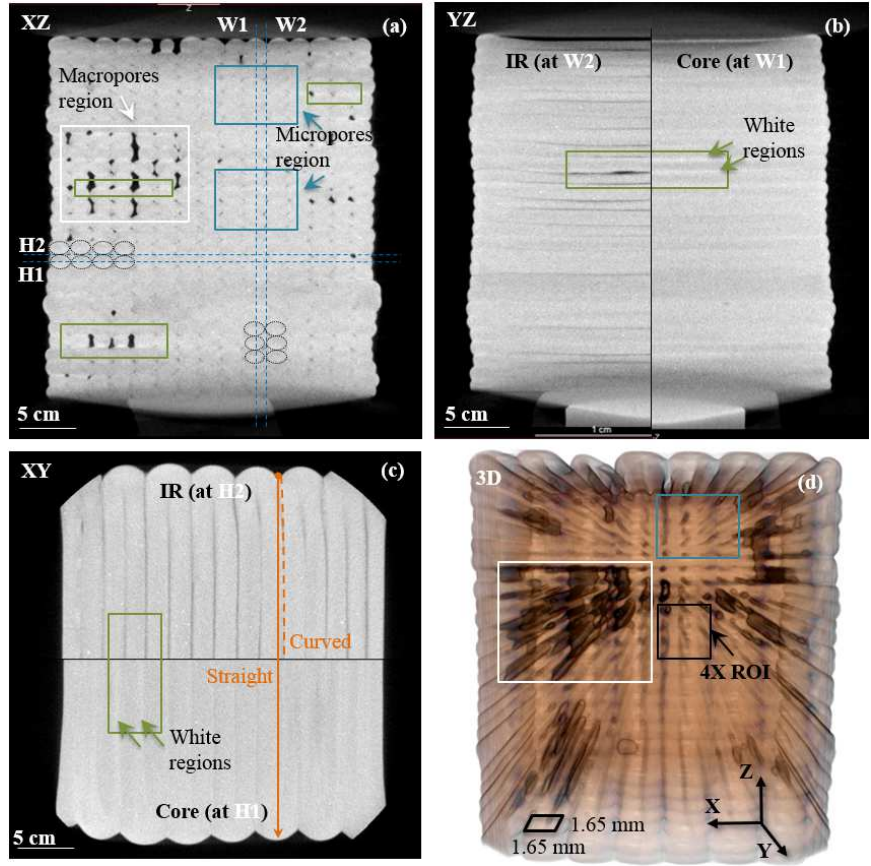

**Fig. 3.** X-ray micro-CT images of the microstructure of 3-days old paste specimen with lamellar architecture collected during 0.4X scan: **a)** 2D projection on the XZ plane; **b)** 2D projection on the YZ plane; **c)** 2D projection on the XY plane, and; **d)** Three-dimensional rendition of the entire volume of the specimen

*Micropores (Micro-channels).* The second type of pores was observed to exist between adjacent filaments of the adjacent layers (see the regions outlined by blue rectangles shown in Figs. 3(a) and 3(d). These micropores (or micro-channels) are smaller than 100 µm in diameter but can be 10s of millimeters long. They are located between filaments along the Y axis (i.e., they are present in the interfacial regions (IRs) of the filaments). The horizontal (i.e., XY) slices H1 and H2 shown in Fig. 3(c), correspond to, respectively, the core (i.e., through the center) and interfacial regions of the filaments as indicated in cross-sectional view (XZ) in Fig. 3(a). Similarly, vertical (YZ) slices W1 and W2 shown in Fig. 3(b) also correspond to the core and interface regions of the filaments indicated in Fig. 3(a). Analysis of Figs. 3(b) and 3(c) indicate that micro-channels (appearing in these Figures as darker regions within the matrix) are only present in the images representing slices through interfaces (i.e. they are absent from images representing slices through the cores). It should be noted, however, that these micro-channels are more pronounced in horizontal than in the vertical planes. This can be clearly seen by comparing images of the IR H2 (shown in Fig. 3(b) and IR W2 (shown in Fig. 3(c)).

Finally, one would expect that the gap formed between four semi-circular filaments will have a diamond-shaped cross-section (when viewed along the axes of the filaments). However, as illustrated in Figs. 4 and 5, this was not the case for the lamellar architecture specimen prepared for this study as the gaps have been found to be triangular in shape. The next sections describe two other characteristics of the printed microstructure (rearrangement of the filaments and accumulation of anhydrous cement grains near macropores), which are considered to be responsible for the formation of these triangular gaps.

*Self-Drifting of Filaments from the Programmed Toolpath.* The examination of CT images revealed that the internal filaments of the lamellar architecture self-drifted (to an extent of about half width of the filament) from their targeted (programmed) toolpath. From the top view (i.e., XY plane), the filaments in all interior layers were observed to follow slightly curved, rather than straight, paths. This is in spite of the fact that the nozzle was programmed to move along a linear tool path parallel to the Y axis. This observation was made possible due to the differences in the gray level intensities between the signals from the core sections and interfaces of the filaments. A typical curvature is highlighted in Fig. 3(c) by the set of two orange lines: the solid line on the left showing the location where a straight interface should have been observed and the dashed line on the right that shows the actual (slightly curved) path of the actual interface. In the consecutive layers of the microstructure, the direction of this curvature successively changes to left and right. This has to do with the opposite direction of printing path in successive layers. The causes of this self-drifting of the filaments are further elaborated in the section describing the 4X scan.

*Accumulation of Anhydrous Cement Grains near Macropores ('White Regions').* The 0.4X CT scans also revealed the accumulation of the unhydrated (anhydrous) cement grains in the IRs near the macropores. This phenomenon was observed in both, the horizontal and vertical interfacial planes of lamellar architecture. In addition, the 4X scan demonstrated the presence of a subtle, 'brighter' zone at the horizontal interfaces, typically also near the macropores. These brighter regions of the microstructure are referred to in this paper as 'white regions' and are outlined by green rectangles in Figs. 3(a), 3(b) and 3(c). The more in-depth analysis of these white regions is presented in the section describing the 4X scan.

**4X Scan.** The 3D scan of the specimens resulting from the 0.4X scan shown in Fig. 3(d) has been used to select an internal region of interest (ROI) to be evaluated at higher (i.e., 4X) magnification. That ROI is outlined by a red square rectangle in Fig. 3(d) and the details are presented in Fig. 4. The examination at the magnification 4X was performed to further explore the four types of microstructural features discussed during the analysis of the results from the 0.4X scan. Specifically, the 4X CT scan provided additional information on the shape of the micro-channels and their connectivity in the IRs, the rearrangement of the paths of the filaments, and the nature of white regions. The 2D projections of individual XZ, YZ, XY planes are shown in, respectively, Figs. 4(a), 4(b) and 4(c); Fig. 4(d) shows the 3D rendition of the interior of the specimen. The higher resolution of these images (compared to the resolution obtained during the 0.4X scans), allowed for the identification of larger cement grains (they appear as bright spots against the gray matrix of the hydrated products) and pores (they appear as dark spots).

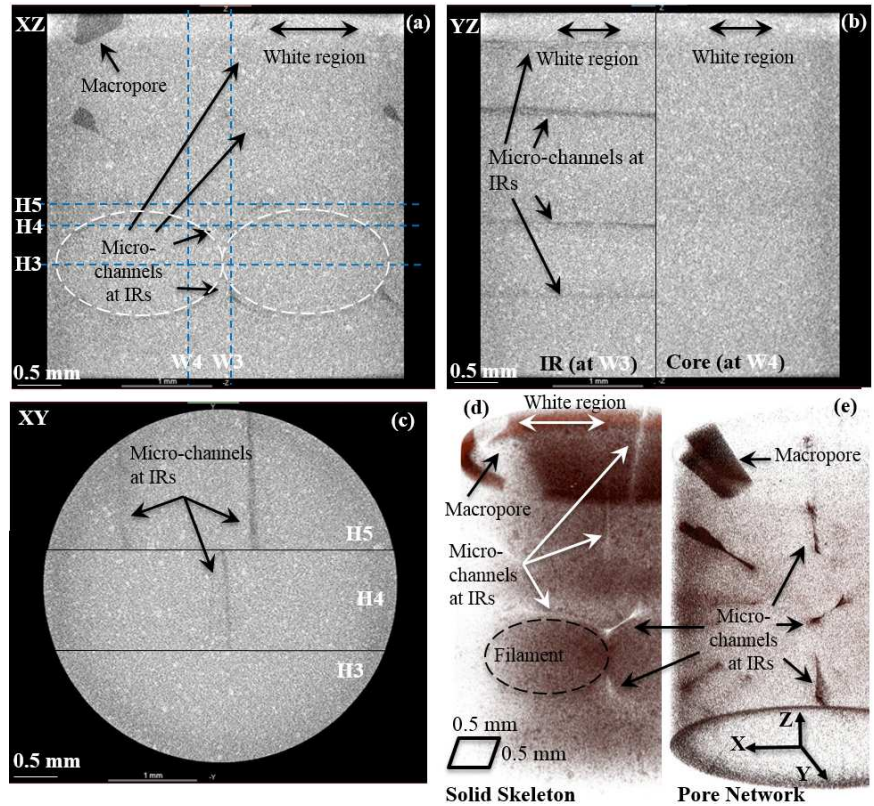

**Fig. 4.** CT images of the microstructure of 3-days old intact specimen with lamellar architecture collected during 4X scan: **a)** XZ; **b)** YZ; **c)** XY planes; **d)** 3D image of the solid skeleton; **e)** 3D image of pore network

*Macropores.* The previously mentioned ROI was selected to capture the bottom edge of a typical macropore between two filaments of the same layer. An example of such pore is shown in the top left corner of Fig. 4(a). This macropore is also shown in the image of the solid skeleton (Fig. 4(d)) and that of the pore network (Fig. 4(e)).

*Micropores (Micro-channels).* This section provides further analysis of the shape of the micro-channels as observed in 4X images. Specifically, it can be observed that the triangular micro-channels ((Fig. 4(a)) are commonly connected to one another and, when viewed in 3D, they form of an inclined dog-bone shape as illustrated in Figs. 4(d) and 4(e). As filaments rearrange upon deposition and move closer together, the connectivity between the micro-channels can be facilitated. Thorough investigation of the 4X scanned volume qualitatively indicated that this connectivity between micro-channels occurs through IRs and form pattern of pore network aligned with the filament architecture (Fig. 4(d) and 4(e)). The typical three horizontal (XY) slices of H3, H4, H5 (indicated in Fig. 4(a)) and shown in Fig. 4(c)), demonstrate the homogeneous characteristics of the microstructure along the 'cores' (H3), compared to heterogeneous characteristic along IRs where micro-channels are present (H4, H5). Similarly, the vertical (YZ) slices of W3 and W4 (indicated in Fig. 4(a)) and shown in Fig. 4(b)), demonstrate the homogeneous characteristic of the microstructure along the 'cores' (W4), compared to the heterogeneous characteristic of the microstructure along 'IRs' (W3).

*Self-Drifting of Filaments from the Programmed Toolpath.* As discussed in the section presenting the 4X scan, the triangular cross-sectional shape of the micro-channels is correlated with the rearrangement of the filaments. This indicates that the filaments are shifted upon deposition towards an adjacent filament deposited before them. This was confirmed via visual observations during printing. Rearrangement of the filament toward a higher packing arrangement shown in Fig 4(e) further confirms this observation. As the filaments rearrange from their designed path, the shape of the pores changes from diamond to triangle as schematically shown in Fig. 5(a) and 5(b). The similarity between the triangular cross-sectional shapes of micro-channels shown in Fig. 4(e) and the illustration presented in Fig. 5(b) confirms that the formation of the triangular-shape pores is the result of self-drifting of the filaments.

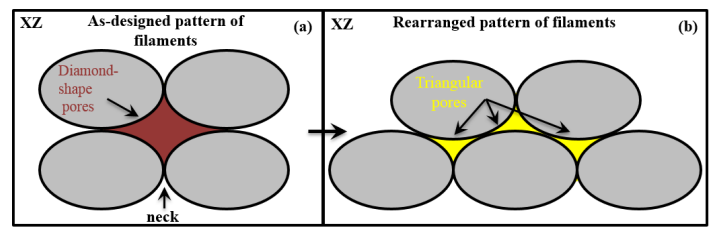

**Fig. 5.** Schematic illustration of arrangements of filaments and shapes of the pores along the Y direction: **a)** As-designed pattern in lamellar architecture containing diamond-shaped pore and; **b)** Rearranged filament pattern showing triangular-shaped pores (similar to those depicted in 4X CT image (Fig. 4 (e))

A variety of mechanisms for interaction and driving force for filament rearrangements can be hypothesized. As an example, during deposition, each filament interacts with the adjacent filament and therefore is shifted towards it. As each filament is deposited, it is designed to overlap and make contact with its adjacent filaments. In addition, the phenomenon such as die swell and relaxation of the filament due to gravity upon deposition may facilitate this contact. Once the filaments make contact, the viscous flow can be driven by the differences in surface curvature between the body of the filament and its neck regions (see Fig. 5(a)), and can bring the adjacent filaments closer. In addition, the rearrangement may cause lamellar architecture to achieve a higher packing density and possibly establish a densifying mechanism of viscoelastic materials upon deposition.

*Accumulation of Anhydrous Cement Grains near Macropores ('White Regions').* The subtle trace of 'white region' at the horizontal interfaces discussed in connection with the 0.4X scan is captured in greater details in 4X (see the top parts of Fig. 4(a) and (b)). A thorough investigation this region demonstrates that the white region contained a higher fraction of anhydrous cement grains, which explains why they appear brighter as cement grains have higher density relative to the density of the hydration products and pores. Similarly, the images of the solid skeleton (shown in Fig. 4(d)) and that of the pore network (shown in Fig. 4(e)), both illustrate the accumulation of anhydrous cement grains in the white region. The fact that white regions are only seen in the horizontal planes suggests that an encouraging drying environment during the printing of each layer occurs in these planes. The presence of the macropores may have additionally enhanced drying. The use of ink with a low water to cement ratio of 0.275 can also be contributing to the increased rate of drying at the horizontal IRs due to longer exposure time to air drying compared to vertical IRs.

### 3.2 Cast Specimen in 0.4X micro-CT (0.4X scan)

**0.4X Scan.** The 0.4X scan of cast specimen showed the random distribution of the pore network. The XY plane of 0.4X scan of intact cast specimen is illustrated in Fig. 6(a) whereas Fig. 6(b) shows the 3D rendition of the interior of the specimen. The darker regions in both images represent pore regions. The random distribution of pores in cast elements and the lack of patterned heterogeneous network compared to those observed in lamellar architecture (Fig. 3), highlights the differences between the microstructure of traditionally cast and 3D-printed hcp specimen.

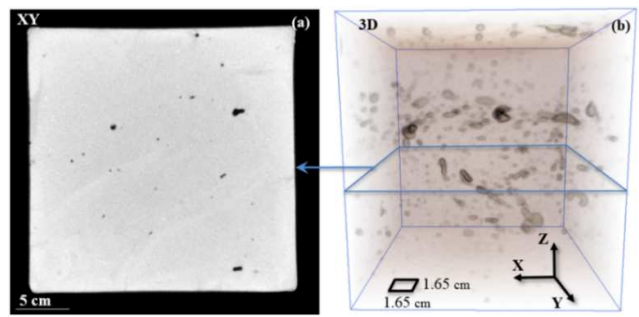

**Fig. 6.** X-ray micro-CT images of the 3-days old intact cast specimen collected during the 0.4X scan: **a)** 2D projection on the XY plane and; **b)** 3D rendition of the entire specimen

**Further Discussion.** The alignment of macropores and micro-channels with the pattern of the filaments in the lamellar architecture (Fig. 3(d) and Fig. 4(e)) suggests that the pore network is inherently associated with the specific architecture of the specimen. However, it is also possible that the location and the amount of the macropores are linked to the specific printing parameters used in this study.

From 0.4X scan, it is evident that the characteristics of the microstructure are different between the solid cores and IRs, induced by the presence of interfacial porosity and the differences in its distribution. It could be hypothesized that the heterogeneous micropore network is present due to the initial presence of extrusion-induced lubricating layer (i.e. layer containing water that surrounds the filament upon deposition) which makes the outer region of the filaments more prone to evaporation thus resulting in the creation of the micropores at IRs. The evaporation and the subsequent wetting at the interface due to deposition of successive layers could also give rise to anisotropic properties of the microstructure and 3D-printed elements. It is also considered that the presence of triangular micro-channels as shown in Fig. 5(b) could be induced during deposition. Given the round shape of the nozzle, air and lubricating water surrounding the filaments can be trapped in between filaments upon deposition of each layer. The presence of lubricating layer followed by drying mechanisms during the print and continuing hydration of cement paste can leave micropores at the IRs across the circumferential zone of the filament as indicated by the dashed oval in Fig. 4(e). From 0.4X scan, it is evident that the characteristics of the microstructure are different between the solid cores and IRs, due to the presence of interfacial porosity and the differences in its distribution. Horizontal interfaces were exposed to air for a longer period of time than vertical interfaces during the printing process. The higher amount of horizontal (dark) IRs compared to (also dark) vertical IRs (Figs 3(c) and Fig. 3(b)) indicates the higher amount of drying and resulting micropores in horizontal planes (XY). This observation indicates the existence of the correlation between exposure time and formation of heterogeneous IRs in different planes, which can result in anisotropy of mechanical properties of the 3D-printed specimen. The accumulation of anhydrous cement grains near macropores in horizontal IRs could be the result of the higher amount of drying at the horizontal IRs. These IRs are shown to be connected to one another through the micro-channels in the microstructure of 3D-printed lamellar architecture as shown in Fig. 4(e). This heterogeneously patterned pore network, together with the presence of macropores, causes anisotropy in the 3D-printed elements. This anisotropy can lead to differences in mechanical properties between the 3D-printed specimen and the cast specimens, depending on the architecture of materials and the interfacial (bond) strength of 3D-printed element in different directions. The presence of the interface is very important as IRs follows the pattern of the filament and overall layered architecture of the element. The architecture can then be designed to allow damage and micro-cracking to be promoted at the weak IR to achieve higher fracture resistance without sacrificing the strength.

It was previously discussed that in 3D-printed hcp elements, a wide variety of heterogeneous features could exist over a broad range of scales [21,22,23]. Application of micro-CT characterization technique for 3D printed lamellar architecture demonstrated the presence of the weak IRs induced by the processing and patterned via architecture. In addition, rearrangement of the filaments and the resulting change in morphology of the pores was revealed. Connectivity of micropores was discovered in the 4X scan in 3D-printed architecture. As revealed by comparing the 4X scans of the printed and cast specimens, their microstructure is very different. These observations reveal the role of processing in determining the microstructure of materials.

## 4    Summary

- Micro-CT images of 3D-printed lamellar architecture collected during this study revealed four characteristic features of the microstructure not observed in the cast specimens, thus indicating these features resulted from the processing of material.
- The previously mention microstructural features included the following: a) macropores (i.e., visible gaps formed during printing), b) micropores at interfacial regions (IRs) of filaments in the form of micro-channels smaller than 100 μm, c) self-rearrangement of filaments from their designed toolpath, and d) high accumulation of anhydrous cement particles near the large pores. All of these four features were qualitatively detected at the 0.4X scans, and further verified at the 4X scan. The presence of these features could play a role in determining the overall mechanical response of architectured hcp specimen.
- Pore network (at both macro and micro scale) appeared to be aligned with respect to the direction of the filaments in the lamellar architecture of the printed specimen.
- Micro-CT demonstrated great aptitude (as non-destructive technique) for capturing spatial heterogeneities of the microstructure of 3D-printed elements as large as 10s of mm. The dual-stage magnifications system used in this study facilitated higher resolutions at large working distance, thus allowing CT of larger specimens.

### Acknowledgements

- The authors gratefully acknowledge generous support from the National Science Foundation (CMMI 1562927) of this research.